# Electricity grid tariffs for electrification in households: Bridging the gap between cross-subsidies and fairness

**Claire-Marie Bergaentzlé** [1*] **Philipp Andreas Gunkel**[1]**, Mohammad Ansarin**[2]**, Yashar Ghiassi-Farrokhfal**[2]**, Henrik Klinge Jacobsen,**[1]

[1] DTU Management, Technical University of Denmark, 2800 Kgs Lyngby, Denmark
[2] Rotterdam School of Management, Erasmus University, 3062PA Rotterdam, Netherlands
* Correspondence: clberg@dtu.dk

**Abstract**

Developing new electricity grid tariffs in the context of household electrification raises old questions about who pays for what and to what extent. When electric vehicles (EVs) and heat pumps (HPs) are owned primarily by households with higher financial status than others, new tariff designs may clash with the economic argument for efficiency and the political arguments for fairness. This article combines tariff design and redistributive mechanisms to strike a balance between time-differentiated signals, revenue stability for the utility, limited grid costs for vulnerable households, and promoting electrification. We simulate the impacts of this combination on 1.4 million Danish households (about 50% of the country's population) and quantify the cross-subsidization effects between groups. With its unique level of detail, this study stresses the spillover effects of tariffs. We show that a subscription-heavy tariff associated with a ToU rate and a low redistribution factor tackles all the above goals.

## 1. Introduction

In the coming decades, massive investments are expected to transform and modernize power grids and achieve national and international goals for a sustainable energy transition. In Europe, projected investments in high voltage grids are estimated at €125-148 billion by 2030 (European Commission 2017) and in distribution grids at €375-425 billion over the same period (Eurelectric 2021a). In Denmark, distribution grid costs may be 20% higher than in the previous decade due to the development of decentralized energy resources (DER) (Hansen, Larsen, and Larsen 2021). This increase raises concerns about how these costs will affect society in general and households in particular.

At the household level, electric vehicles (EVs) and heat pumps (HPs) are widely considered to be key drivers of $CO_2$ emissions reductions. Projections indicate that 45 million electric heat pumps and 50 to 70 million electric vehicles will be connected to European grids by 2030 (Eurelectric 2021b). However, the widespread adoption of EVs and HPs also raises three interrelated issues. First, these technologies have relatively high investment costs, favoring access to the wealthiest households, despite existing subsidies. Second, electrification of transportation and heating increases electricity demand and peak effects, which increase system cost. Finally, the flexibility of demand brought by car batteries, and to a lesser extent by HPs, may lead to load deferral. This situation can shift the recovery of grid costs to grid users who do not have access to these technologies. Hence, the market uptake of EV and HP in the context of heavy network investments will place an additional burden on less flexible households, including the poorer ones. However, to what extent this change might affect low-financial status or passive households remains overlooked.

This article addresses three research gaps. A growing literature is emerging to understand better the effects of tariffs on the activation of flexibility, stranded grid cost recovery, and fairness in society. To the best of our knowledge, little research has developed and compared multiple tariff designs based on these three dimensions simultaneously. Besides, most past studies focus on case studies highlighting tariff impacts (flexibility enabling, grid bill saving, etc.) at the scale of a typical household. Including the fairness dimension implies widening the scope of the analysis to the societal scale and thus capturing the effects of tariff cross-subsidization between categories of grid users showing different socio-economic and technology (EV, HP) ownership characteristics and load patterns. To our knowledge, there is no comprehensive study quantifying the unfairness effect of tariffs with the adoption of HP and EV in a broad population. Finally, a branch of the literature investigates how redistributive mechanisms can redistribute costs across actors. In this study, we propose and test an original redistribution instrument embedded in the tariff that ensures the fair allocation of grid costs across the grid users. Ultimately, the tested tariffs and redistributive instruments open new doors for sending the right economic signals for grid cost recovery and flexibility, supporting household electrification, and "leaving no one behind", to borrow the phrase from the European Green Deal (The European Commission 2021).

We simulate the impacts of five two-part tariff designs on 1.4 million Danish households (about 50% of the country's population). We measure in detail and comprehensively present the cross-subsidization effects of each tariff on household categories, defined as low, medium, and high-financial status, with or without EV or HP. We also include in the analysis a redistribution factor that alleviates part or all of the subscription payment faced by low financial status households. Our results give insights into the cross-subsidies effects across households and point to potentially unexpected spillovers of a rate design or redistributive policies on specific households. We inform decision-makers on how to strike a balance between the conflicting targets of more cost-reflective grid tariffs and fairness in the context of electrification. We also offer a reflection on the current state of regulation limiting interactions between rate making and energy policy. It is out of the scope of this study to determine which of the simulated tariffs is the most effective in improving social welfare. Instead, our results inform policy-makers, system operators, and society of the possible consequences for consumers associated with tariff designs in the context of electrification. We leave it to policy-makers to engage in a normative reading of our results. The developed method allows others to perform similar analyses on grid user groups, including non-residential.

In section 2 of this article, we present the recent literature covering grid tariff designs for flexibility, revenue adequacy and fairness and present out-of-tariff redistributive mechanisms for fairness. Section 3 outlines the data and households' categorization. Section 4 presents the tested tariff scenarios. Section 5 defines our method, Section 6 displays the results. Section 7 offers a policy discussion, and Section 8 concludes.

## 2. Literature review

Efficient electricity pricing is historically considered as sending appropriate signals for allocative efficiency (Farrell 1957). In his classic work on utility rates, (Bonbright 1961) sets out three general principles to guide tariff design for public goods based on i) short- and long-term economic efficiency, ii), revenue adequacy and iii) fairness between users. Given the nature of electricity as a network good (Heald 1997) and the high proportion of final costs associated with grid development (Simshauser 2016), electricity pricing is particularly driven by these principles.

The first principle links allocative and productive efficiency to cost-reflectiveness. This principle is theoretically reached when pricing electricity based on consumers' inverse price elasticity, as shown in (Boiteux 1951). In this setting, a cross-subsidy effect implicitly exists between flexible and inflexible users, with the latter ultimately paying for the fixed network costs, which is unpopular for public goods pricing (Bonbright 1961; Neuteleers, Mulder, and Hindriks 2017). The second implies that marginal cost pricing, or first-best pricing, ineluctably leads to bankruptcy due to networks' increasing returns to scale. The third principle establishes that electricity costs should be apportioned across consumers depending on the relative burden their use incurs to the system. From the public-sector perspective of many regions, including Europe, this issue of fairness is a significant concern. This concern is gaining attention within the European Green Deal via the context of "fair transition". Here, points of attention (and contention) are challenges in the fair social distribution of the costs of the sustainable transition. These three principles summarize the dilemmas of network pricing with electrification and DER growth. In this context, the challenge is finding the right balance between cost-reflective signals, revenue adequacy for the utility, and cross-subsidization effects between consumers presenting different load profiles.

In addition to economic concerns, electricity grid tariffs have also been historically subject to political considerations. Tariffs have for example been used as instruments for national socioeconomic development and the promotion of the economic inclusion of society's vulnerable (Yakubovich, Granovetter, and Mcguire 2005; Heald 1997). Often, these goals pull tariff design away from the economically best-case solution, see e.g. the unpopularity of first-best (Ramsey) pricing in residential settings (Neuteleers, Mulder, and Hindriks 2017) and that in no case had the public authorities (at the time of writing of (Boiteux 1956)diverged from a stable tariff. Yet, the progress made in smart grid technologies enables the generalization of dynamic price signals, including grid tariffs. Increasingly, rate making bridges the gap between the principles of economic efficiency and cost recovery. However, the (un)fairness of new tariffs remains unclear.

On the tariff-economic efficiency nexus, cost-reflective tariffs build on electricity's peak demand periods in introducing time-differentiated rates, either in a fixed and simple way with Time-of-Use tariffs which place higher kW/h rates during periods of larger consumption, or in more specific ways, e.g. with critical peak pricing schemes (A Faruqui and Palmer 2012). Other tariff types use a capacity component. Non-coincident demand charges (based on the peak demand of a single household over a given period) have been considered. However, they have questionable cost reflectivity due to pricing a peak different from the network peak (Borenstein 2016). This is often fixed in the second proposed option, coincident demand charges (based on household contribution to network peak demand over a long time period), which can be unpredictable and uncontrollable for households (Burger et al. 2019; Simshauser 2016).

On the tariff-revenue adequacy nexus, a simple solution is sometimes offered as increasing fixed charges, in two-part tariffs, which do not depend on consumption (Schittekatte, Momber, and Meeus 2018; Bergaentzlé et al. 2019; Bergaentzle and Gunkel 2022). Multiple utilities increase the subscription part to move this payment closer to fixed cost (Faruqui 2021), to prevent financial imbalance for the grid operator, and limit cross-subsidies across grid users (Clastres et al. 2019). This latter type of tariff that builds on a strong fixed component and a volumetric component, which itself can be established in ToU tends to strike a balance between more cost-reflectiveness and limited financial risk for the utility.

The tariff-fairness nexus has resurged interest with the rapid growth of DER, EVs, and HPs (Ansarin et al. 2022; Lamb et al. 2020; Pollitt 2018). This interest has been particularly aimed at the residential sector, with the goal of remediating some of the largest inequalities resulting from environmental policies supporting energy technologies (Peter Grösche and Schröder 2014). Regarding technologies, the fairness issues of installation costs and feed-in tariffs of solar photovoltaic panels is an especially well-studied subject (Winter and Schlesewsky 2019; Farrell and Lyons 2015; Nelson, Simshauser, and Nelson 2012; Nelson, Simshauser, and Kelley 2011). Many papers in this context report on policy options for blunting their inequities (Burger et al. 2020; Borenstein 2012). Reviews of this subject can be found in (Ansarin et al. 2022) and (Lamb et al. 2020). Recent literature also documents that the uptake of DER and EVs with unsuitable tariffs causes revenue instability for system operators, potentially resulting in revenue inadequacy. This risk further shifts the payment of network costs to users who do not have the possibility of self-consumption and flexibility (Abdelmotteleb et al. 2018; Barbose and Satchwell 2020; Haro et al. 2017; Hoarau and Perez 2019; Neuteleers, Mulder, and Hindriks 2017; McLaren et al. 2015). As with DER, EVs and HPs have had rapid growth in many regions and have significant influence on household demand profiles. Nonetheless, less attention has been given to the fairness issues caused by these technologies in different tariff scenarios.

The fairness concern has been defined differently in different contexts by different stakeholders. (Burger et al. 2019) discusses three competing definitions, of which we use the definition of "Allocative Equity". An allocatively equitable tariff treats identical customers equitably. This definition corresponds to those commonly used in economics literature (Reneses and Ortega 2014; Bonbright 1961), where tariffs should fairly apportion costs among different grid users and avoid undue discrimination. Under the assumption that any change in tariff does not affect the overall revenue for a utility (revenue neutrality principle), any allocative inequity would be transferred between customers. Thus, this inequity is the support of some customers' consumption by other customers, namely cross-subsidization. Accordingly, a branch of research also focuses on non-tariff-based redistributive policies for electricity pricing unfairness. Such redistributive policies complement tariff design if the tariff hinders overarching policy targets (e.g. guaranteeing a fair price of energy or limiting energy poverty). Several redistributive policies support utility payments for low-income energy consumers through tax-based money transfers. In the residential sector, the most widely used and considered policies include means-based direct funds transfers (Borenstein 2012; Neuhoff et al. 2013; Frondel, Sommer, and Vance 2015), progressive fixed charges (Burger et al. 2020), energy efficiency-supporting measures (Neuhoff et al. 2013), or income payment plans. The latter is especially gaining attention in the US. In Ohio, a percentage of income payment plan is running to lower poor households' utility bills through transfers from wealthier households. A similar programme is under review in Knoxville (Voice, n.d.; The Advertiser Tribune, n.d.).

Redistributive factors are another instrument especially used for tariff-based money transfers. These factors are designed to ensure that households living in vulnerable financial situations have access to some electricity (Barbose et al. 2021; Borenstein 2012). This policy commonly requires that other households within the distribution grid pay a surcharge to support this expense (Burger et al. 2020). In Europe, the EU Regulation 2019/943 Article 18 (European Parliament 2019) confines the scope of action of utility regulators for rate-making to an obligation to certify the nondiscriminatory and cost-reflectiveness requirement in tariffs and dissociates tariff-making with other political goals such as fairness. This also applies to the EU country, Denmark, in being transposed into the electricity supply law (elforsyningsloven), §73 (Folketinget 1999). However, the governments of many European countries intend to pursue policies of electrification and consumer protection from extreme rises in energy prices. Hence, it is necessary to explore all policy options available for striking a balance between economic signal, decarbonization and fairness in electricity grid pricing.

## 3. Data and household categorization

This section describes our data, our grouping of households based on socio-economic attributes, and their categorization into low, medium, and high financial status baskets.

### *3.1. Dataset – coupling hourly-metered data with socioeconomic data*

We followed a three-step approach for data collection and cross-featuring. First, we collected anonymized data from Statistics Denmark and cross-featured it to national socioeconomic registers on citizens, families, households, vehicles, buildings, and addresses. All the collected data are from 2017 registers. The raw data covers approximately 4 million out of Denmark's 5.6 million citizens, representing about 2 million households. Second, we cross-featured the households with hourly electricity metering data collected from the Danish EnergiHub operated by the Danish TSO, Energinet. 1.5 million smart meter ID points were available for the full year of 2017, when smart meters roll-out was still ongoing. Each meter ID gives access to hourly-metered electricity use. Finally, we removed from the final dataset all the entries showing disturbances in individual load profiles resulting from a malfunction of the meter or a late installation. All metered data showing faulty entries greater than a thousand hours (of 8760 hours in the whole year) were first excluded, then rebuilt using the average consumption value for the same household category, as defined in the next section[1]. In the end, this study covers approximately 1.4 million households.

*3.2. Household's socio-economic attributes*

Five socio-economic attributes serve the consumers' grouping (Table 1). Dwelling types are divided into houses (denoted by *H*) and apartments (*AP)*. Dwelling areas are divided into different sizes based on median statistics of dwelling areas. Since houses and apartments have different size characteristics, house sizes are categorized into *A1* (up to 110 sqm), *A2* (110 to 146 sqm), and *A3* (over 146 sqm), whereas Apartment sizes are classified as *A1* (up to 66 sqm), *A2* (66 to 85 sqm), and *A3* (over 85sqm). Occupancy gives the number of persons living in the household. P1 and P2 indicate respectively one and two residents, P3+ contains households with three or four occupants, P5+ represents five or more occupants. Household annual gross income is used to categorize three groups based on median statistics. The lower third income group, represented by *€1*, earns up to 240,260 DKK/year, the medium income group *€2* up to 449,097 DKK/year, and the remaining upper group is included in *€3*. Lastly, households are separated based on whether they own EVs (*EV1*) or not (*EV0*) and whether they own HPs[2] (*HP1*) or not (*HP0*).

---

[1] Note that replacing some timeslots with averages leads to close-to-average calculations for costs, and thus leads to a conservative estimate of unfairness amounts.

*Table 1: Attributes used in households categorization*

| Type of dwelling | |
|---|---|
| House (H) | Apartment (Apt) |
| **Size of dwelling (A)** | |
| A1: A1<110m2<br>A2: 110m2<A2<146m2<br>A3: 146m2<A3 | A1: A1<66m2<br>A2: 66m2<A2<85m2<br>A3: 85m2<A3 |
| **Occupancy (P)** | |
| P1: 1 person;<br>P2: 2 people<br>P3+: 3 to 4 people<br>P5+: 5 and more people | |
| **Income level (€)** | |
| €1: <240kDKK<br>€2: 240kDKK<€2<449kDKK<br>€3: 449kDKK<€3 | |
| **Advanced equipment** | |
| HP0: No heat pump<br>HP1: With heat pump<br>EV0: No electric vehicle<br>EV1: With electric vehicle | |

Note: 1 Danish Krone (DKK) = 0,13 Eur. or 0,15 Dollars

All possible combinations of characteristics form 210 unique groups. 120 groups do not include enough households to fulfil privacy regulations and are thus excluded. For a similar reason, households owning both EVs and HPs are also excluded. Due to their negligible size, these exclusions are not expected to materially impact the study's quantitative results. We use the remaining 90 groups in this study.

---

[2] Our analysis includes only HPs from the various options for heating. Other electricity-to-heat options are less commonly used as the primary heating source in Denmark.

***3.3. Categorization of households between high, medium and low financial status***

Prior studies determined that financial factors play a significant role in adopting energy technologies (De Groote and Verboven 2019; De Groote, Pepermans, and Verboven 2016; Gautier and Jacqmin 2020). (Barbose et al. 2021) considers that wealth and income are key determinants for household decisions on solar photovoltaic panel installations in the US. Similarly, both factors would directly influence household purchase decisions regarding HPs and EVs. Hence, we consider a composite measure of both income and wealth in the grouping of households by financial status.

We use income values directly from the dataset and utilize dwelling area as a proxy for wealth. The final categorization contains in total 8 different household groups summarized in Table 2 and detailed Appendix 1. No Tech refers to households that own no HPs or EVs. We consider the households that combine the smallest dwelling area and lowest income group as households with low financial status, regardless of the level of occupancy. Households of this type owning a HP and of up to 2 people are also considered low financial status, as this type of setup is often associated with social housing in Denmark. Medium financial status households include households belonging to the high income group with medium and small dwelling area (house and apartment) without advanced equipment, with HP or with EV. They also include households from the medium and the small-income group with all dwelling area (house and apartment) without advanced equipment or with HP. High financial status households have both the largest dwelling area and highest income, without technology or owning an HP or an EV.

*Table 2: Summary of the studied household groups*

| Financial status level | Advanced equipment |
|---|---|
| Low | No Tech |
| Low | HP |
| Medium | No Tech |
| Medium | HP |
| Medium | EV |
| High | No Tech |
| High | HP |
| High | EV |

## 4. Tariff scenarios and redistribution factor

### 4.1. Base case tariff scenario and considered grid cost

In 2017, Denmark used a binomial tariff (CEER 2017). The fixed per-unit charge of 18.25 øre/kWh (2.43 c€/kWh) (TREFOR 2017) was combined with a fixed annual subscription charge. Thus, the majority of households in our data set present a 2017 demand that was likely not responding to grid tariff price signals. This study refers to this fixed volumetric tariff as the base case.

The aggregated revenue generated by all the tested tariffs must equal the revenue derived from the base case (i.e. tariffs are revenue-neutral). We consider a proportion of total Danish network cost based on the households included in the dataset. For the volumetric component of the tariff, we multiply the volume consumed in our dataset (4150 GWh) by the per-unit fee (18.25 øre/kWh). For the fixed subscription component, (Dansk Energi 2019) estimates an average payment of 428.8 DKK/year for a representative Danish household consuming 4,000 kWh per year. These two components are summed, and the calculations are displayed in Table 3. The resulting total network cost is 1.39M DKK/year, of which 55% is covered by the volumetric part and 45% by the subscription charge.

*Table 3: Breakdown of total network costs and recovery*

| Total consumption in 2017 (GWh) | 4,150 | |
|---|---|---|
| Price (øre/kWh) | 18.25 | |
| Subscription payment per household (DKK) | 428.8 | |
| Total number of households | 1,468,686 | |
| Total cost recovered via the volumetric part (DKK) | 757,409,794 | 55% |
| Total subscription cost (DKK) | 629,772,557 | 45% |
| **Total distribution network cost for 2017 (DKK)** | **1,387,182,351** | **100%** |

## 4.2. Presentation of the tariff designs

In 2020, Denmark introduced a binomial ToU tariff, using a similar subscription fee and differentiating prices in the volumetric part according to the time of day and season (TREFOR 2022). We build on this new tariff design and vary the relative share of the subscription and volumetric part and of the peak and off-peak block rates. We use a three-step approach to design the tariffs (Figure 1).

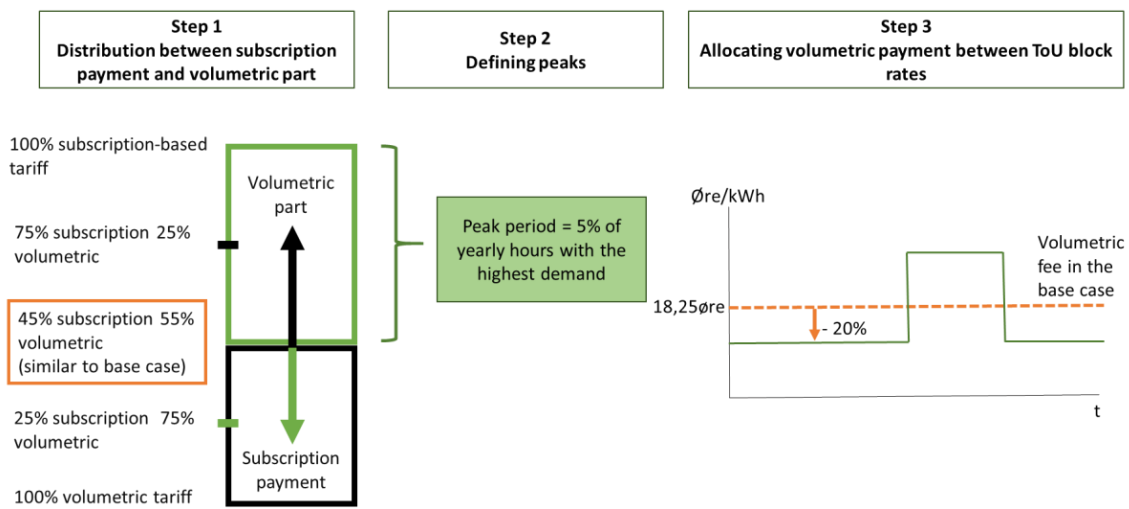

*Figure 1: Steps and parameters used for tariff design*

First, we consider besides the base case five tariff designs corresponding to changes in the relative share of the fixed and volumetric component (Figure 2): a 100% volumetric tariff (and no subscription charge); a 25% fixed part / 75% volumetric part; a 45% fixed part / 55% volumetric part; a 75% fixed part / 25% volumetric part; and a 100% subscription-based tariff (and no volumetric part). In the latter case, all grid users recover the total network cost equally. The 45%/55% scenario is a direct comparison with the base case scenario, with the only modification coming from introducing a peak and off-peak rate in the volumetric part.

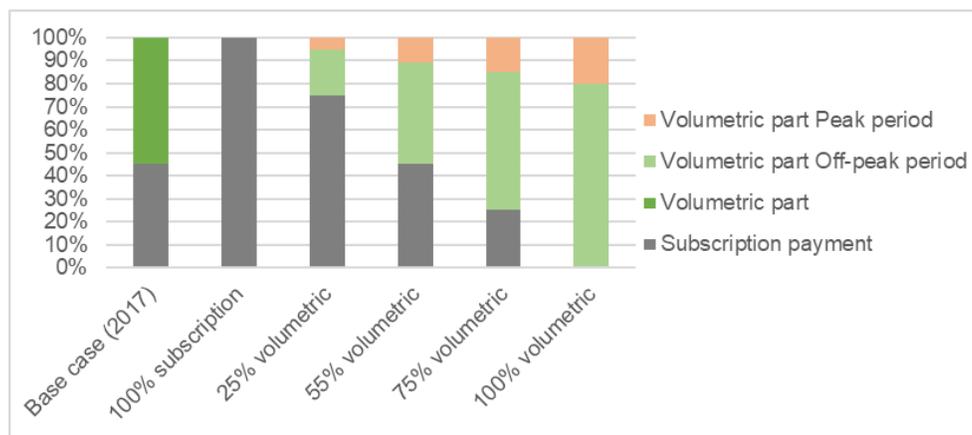

*Figure 2: Illustration of the simulated tariffs.*

Second, we define what is considered a peak period to build the peak rates in the volumetric part using the 2017 Danish load duration curve as a proxy for hours with network congestion. We build on (Ahmad Faruqui et al. 2007) and consider a peak period as 5% of yearly hours with the highest demand.

Third, we set the recovery factor of 0.8 to determine the share of the network cost that will be recovered during the peak and off-peak block rate when a volumetric part applies. This assumption follows the stepwise policy approach by Danish regulators toward off-peak charges reflecting operational cost and peak charges covering long-term investment (European Commission 2015; NordREG 2021) and is validated through approximation with newly implemented grid tariff levels of the Danish distribution system operator (DSO), Trefor (TREFOR 2022). In the 100% volumetric case, that means that a per-unit fee of 14.6 Øre/kWh applies to consumption during off-peak hours (95% of the year) and a per-unit fee of 66.52 Øre/kWh during peak hours (5% of the year). The tariff value of each block in the ToU is therefore determined by the model in each scenario based on the relative share of each cost component and under the constraint of revenue neutrality.

### *4.3. Redistribution factors*

In this study, we test the impact of 10 redistribution factors that allow socially progressive compensation for the high burden of subscription payments. We define the redistribution factor as a portion of the fixed subscription payment that is transferred in steps of 10%, from 0% to 100%, away from low financial status households (and is added to the subscription payment paid by high and medium financial status groups). A distribution factor of 1 indicates that no such transfers are made (no redistributive policy), while a factor of 0 indicates that the other groups entirely cover the subscription payment of low financial status households.

### *4.4. Calculation of equity*

We focus on one measure to compare the equity among different tariff setups. This measure represents the difference in household costs between the base case and the other tariffs as a percentage of payments in the base case. This represents how much each household category may be expected to pay more or less compared to what was paid with the original tariff. Assuming revenue neutrality, increasing payments for some households can be indirectly attributed to cost decreases for other households. Hence, Similar to (Horowitz and Lave 2014; Burger et al. 2019; Simshauser and Downer 2016), this measure can represent the (in)equity of each tariff compared to the base case.

## 5. Method

We simulates the impacts of the tariff designs under the constraint of revenue neutrality for the utility. The model allocates the total expected grid revenue (Table 3) in the different scenarios across grid users and compares the relative share of the total network cost covered by the different groups. Equation (1) satisfies revenue neutrality throughout the different tariff scenarios.

$$R^{SO} = \sum_{g}^{G} (C^{Subsr,init} + C^{Subsr,var}) + f^{Vol,reduc}(q_g^{peak,year} gt^{peak} + q_g^{base,year} gt^{base}) \quad (1)$$

$R^{SO}$ is the total income recovered by the utility through the tariffs. The left-hand side of the sum represents the subscription payment. $C^{Subsr,init}$ is the initial subscription fee paid in the base case. The initial subscription fee is equal for all households and its sum corresponds to the total subscription cost in Table 3. $C^{Subsr,var}$ is the only variable in the equation. It adjusts the subscription fee in response to the respective share of the subscription and volumetric part implemented in the scenarios. Both terms are scalars and independent of the group. The right-hand side of the equation is the volumetric component consisting of a two-block ToU. $q_g^{year}$ is the yearly consumption per consumer category further subdivided into peak and base consumption, respectively $q_g^{peak,year}$ and $q_g^{base,year}$, and multiplied by its respective base and peak tariff value: $gt^{peak}$ and $gt^{base}$. The entire volumetric part is multiplied by the reduction factor $f^{Vol,reduc}$ that adjusts the total revenue recovered by the volumetric part depending on the tariff scenarios. We use a stepwise reduction in $f^{Vol,red}$ corresponding to the five relative shares of the volumetric part ([100%...0%]) tested in the tariff scenarios.

By default, equation (1) calculates the tariff scenarios with a redistribution factor of 1, i.e., without redistribution across groups. Equation (2) introduces the redistribution factors presented in 4.3.

$$R^{SO} = \sum_g^G f_g^{redistr,var}(C^{Subsr,init} + C^{Subsr,fix}) + f^{Vol,reduc}(q_g^{peak,year} g^{peak} + q_g^{base,year} gt^{base})$$

(2)

The previous variable $C^{Subsr,var}$ becomes a fixed scalar $C^{Subsr,fix}$. The value equals the different levels of $C^{Subsr,var}$ depending on the respective scenario of $f^{Vol,reduc}$ calculated in (1). The redistribution factor $f_g^{redistr,var}$ sets the subscription level from the low financial status group such that will be cross-subsidized by the other groups from We use income values directly from the dataset and utilize dwelling area as a proxy for wealth. The final categorization contains in total 8 different household groups summarized in Table 2 and detailed Appendix 1. No Tech refers to households that own no HPs or EVs. We consider the households that combine the smallest dwelling area and lowest income group as households with low financial status, regardless of the level of occupancy. Households of this type owning a HP and of up to 2 people are also considered low financial status, as this type of setup is often associated with social housing in Denmark. Medium financial status households include households belonging to the high income group with medium and small dwelling area (house and apartment) without advanced equipment, with HP or with EV. They also include households from the medium and the small-income group with all dwelling area (house and apartment) without advanced equipment or with HP. High financial status households have both the largest dwelling area and highest income, without technology or owning an HP or an EV.

Table 2. At factor 0, all low financial status households $g$ have a 0 entry, meaning that those households are effectively excluded from the subscription fee. The residual households have the same variable $x^{incr}$ as a variable increasing value (>1) to subsidize it and to satisfy equation (2) and revenue neutrality. $f_g^{redistr,var}$ is thus a vector consisting of 0 and $x^{incr}$ and solving $x^{incr}$ yields the redistributed yearly network cost of every group.

## 6. Results

### 6.1. Final impact on households' grid bill

Figure 3, detailed in Appendix 2, illustrates how the base case and each tested tariff affect the average annual payment by household type under the two extreme redistribution factors: factor 1, without redistributive policy and factor 0, full redistribution.

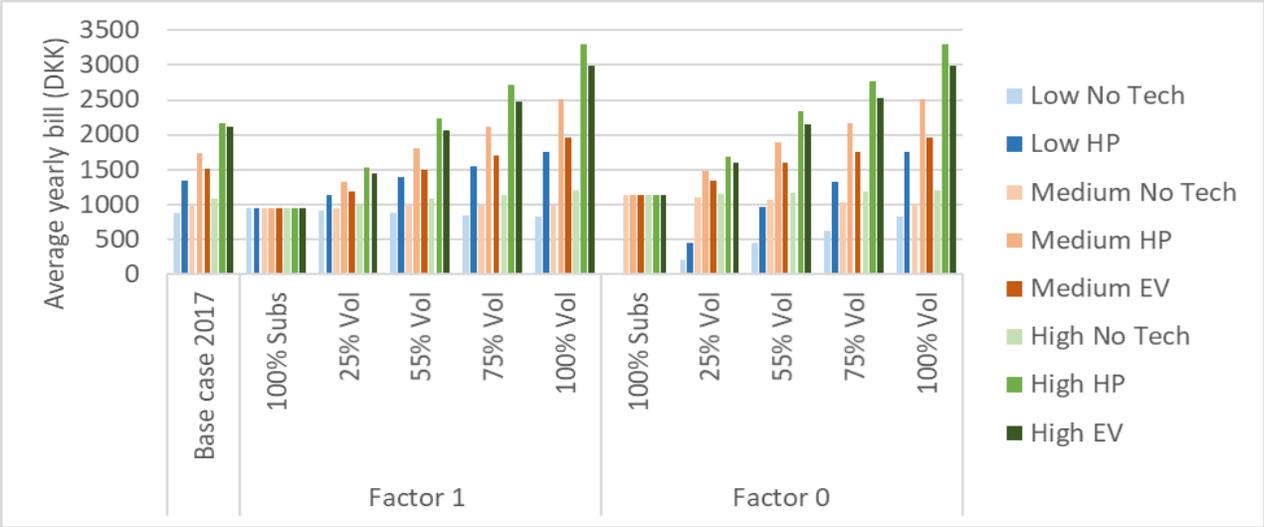

*Figure 3. Relative share of each tariff cost component in the average yearly bill per user category under factors 1 (no redistributive policy) and 0 (full redistribution of the subscription charge).*

Compared to the base case, the new rates yield two main results. First, the relative share between subscription and volumetric components has an inverse effect on low financial status households on the one hand and medium and high financial status households on the other without advanced technology. Low financial status households benefit most from a high volumetric component. A 100% volumetric rate results in an average annual network bill savings of 6% for this user category compared to the base case, while a 100% subscription rate increases the same bill by 7%. On the other hand, medium and high financial status households end up increasing their annual bill by 11% under a 100% volumetric tariff and reduce it by 13% with a subscription-based rate, compared to the base case.

Second, subgroups with EV or HP experience a significant reduction in their average annual tariff cost when the subscription rate increases. The 100% subscription tariff involves that total network costs are equally shared among households. In particular, high financial status households with above-average consumption, such as EV and HP owners, benefit most from it as their grid bill more than halves compared to the base case (-57%). Medium financial status households reduce their bill by up to 46% and low financial status households by 30%. On the contrary, a 100% volumetric rate increases the annual bill of EV and HP owners by 32% to 52%, depending on the financial status, HP owners suffer the most severe increase of their yearly bills (up to +52%) given the continuous consumption profile of heat pumps.

Our results are specific to the relative share of users in each category, which is subject to dynamics. The impact on the grid bill per user will change as more consumers gain access to EVs and HPs, or along with changes in their financial status. Nevertheless, the results provide insights into the winners and losers in each tariff setup, as further detailed below.

*6.2. Relative grid cost recovery per cost component and user group*

Figure 4, detailed in Appendix, shows the average cumulative annual grid bill per user category and illustrates the weight of each tariff component in the two extreme redistribution scenarios.

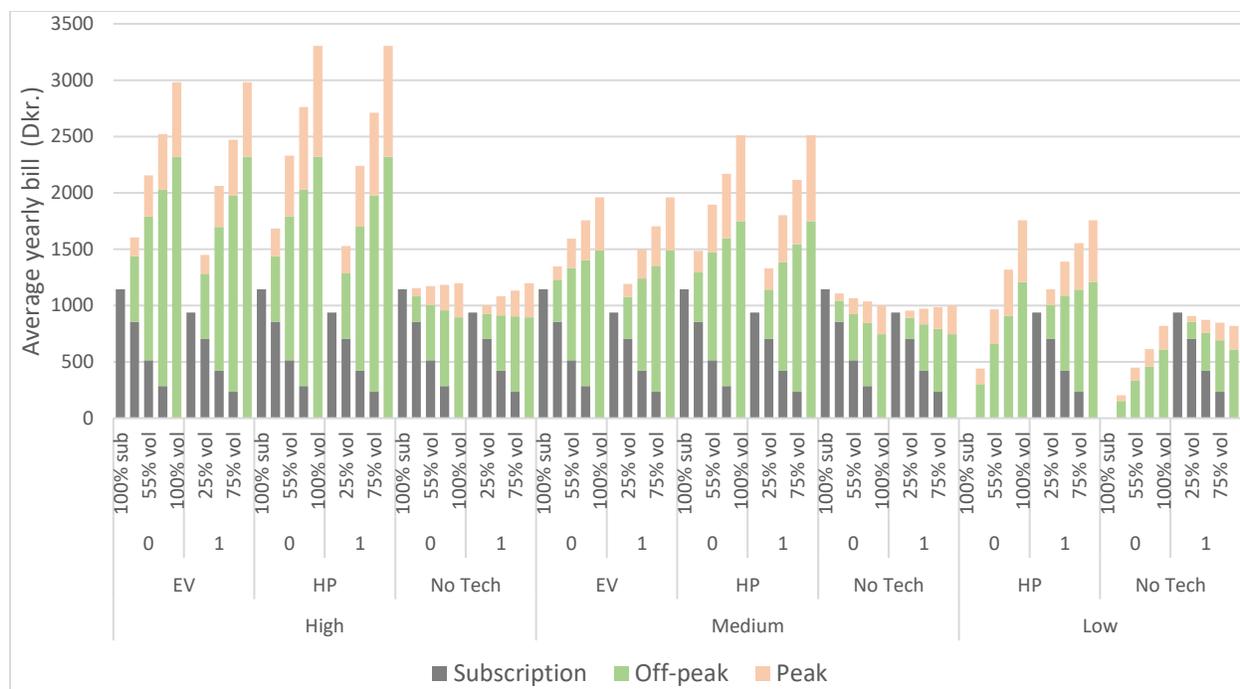

*Figure 4: Relative weight of the cost components per user group in each tariff under factor 1 and 0*

In absolute terms, only the payment related to the subscription part shows differences between redistribution factors 0 and 1. Low financial status households pay no or less subscription charges depending on the tariff design, as medium and high financial status households pay all to part of their payment with the introduction of the redistribution factors. Payments for the peak and off-peak components are unchanged when comparing the same household type and tariff scenario, regardless of the redistribution factor.

Figure 5 captures to what extent the highest factor (factor 0) affects the average grid bill at the household type level. In the 100% subscription tariff, a household with a low financial status avoids an annual expenditure of 938 DKK. (Eur. 126) and other household types would have to pay a surcharge of 205 DKK. (Eur. 27). Since the factor's effect is linear, the results indicate that each additional DKK spent by medium and high financial status households avoids a 4.5DKK spending in the low financial status category.

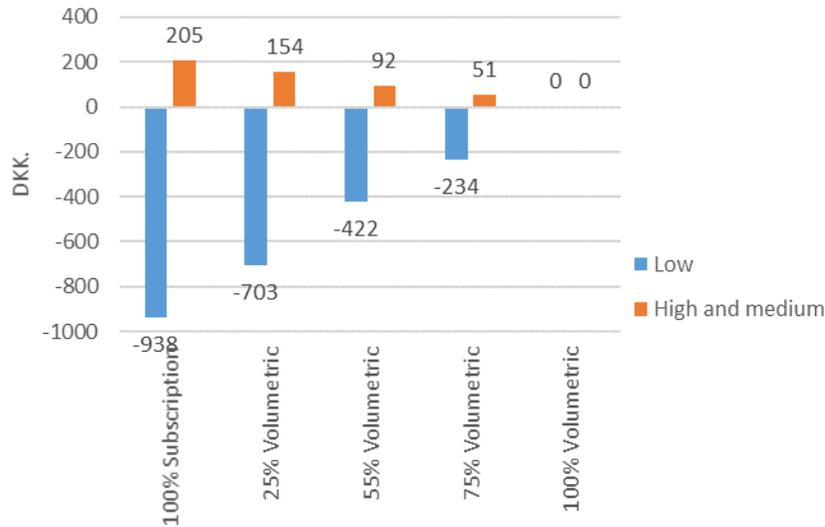

*Figure 5: Impact of the redistribution factor on the grid bill paid by an average household with low financial status and medium and high financial status*

This transfer is relative to the number of households associated with the user category being subsidized and the subsidizing user category. In this case study, we consider ≈18% of the Danish population with low financial status. The smaller the number of households benefiting from the transfer, the smaller the burden on the rest of the population.

### 6.3. Zoom on the redistribution factor

Figure 6, Figure 7, and Figure 8, detailed in Appendix 4, show the difference compared to the base case in the yearly network bill paid by each household group under each tariff setup and all redistribution factors. The difference in payments is calculated based on what each category pays under the base case, as detailed in Appendix 5.

The effect of the redistribution factor is seen in the opposite direction of the different slopes between low financial status households and high and medium financial status households. As the cross-subsidization increases (factor 1 to 0), the relative savings of high and medium financial status households relative to the base case decrease and inversely for the low financial status households. This effect is linear with the progressive introduction of redistribution policy and shows a constant and predictable change in the amount paid relative to the subscription.

For the low financial status households (Figure 6), the slope of this effect decreases as the volumetric share increases because the redistribution applies to a smaller portion of the tariff. This change is steeper for the group without technology than for the group with HP, showing that consumers with the lowest electricity demand benefit from the largest bill savings along with a higher redistribution factor.

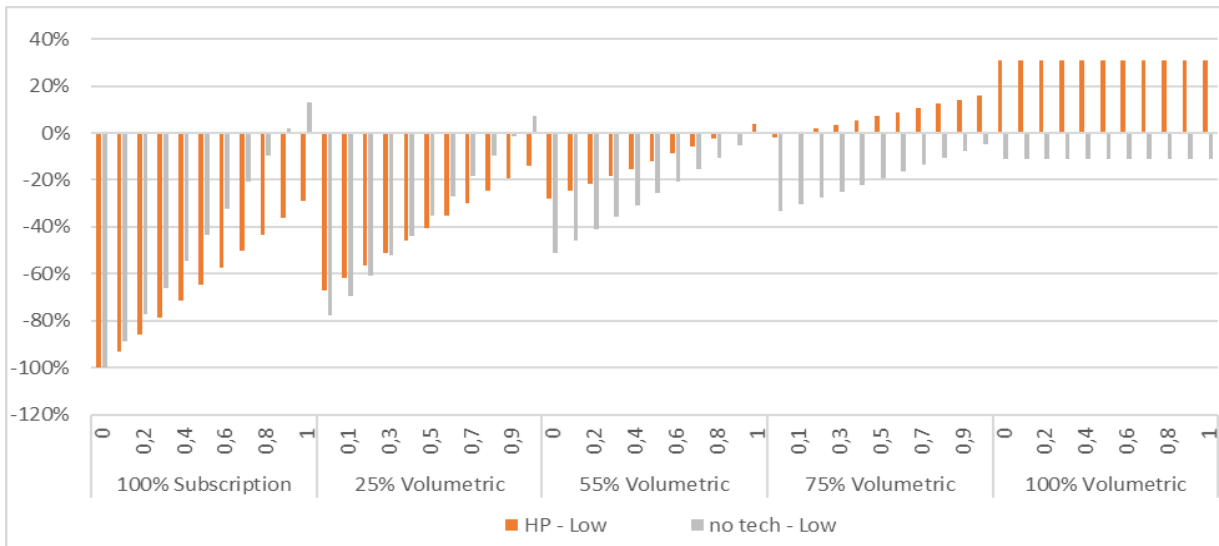

*Figure 6: Impacts of the redistribution factor for each tariff design and consumer group, with and without advanced equipment for low financial status households*

Zooming in on the medium and high financial status users (Figure 7, Figure 8), we see that the redistribution factor generally induces a progressive decrease of their savings relative to the base case (from 100% subscription to 25% volumetric), or a progressive increase of their additional cost (from 55% to 100% volumetric).

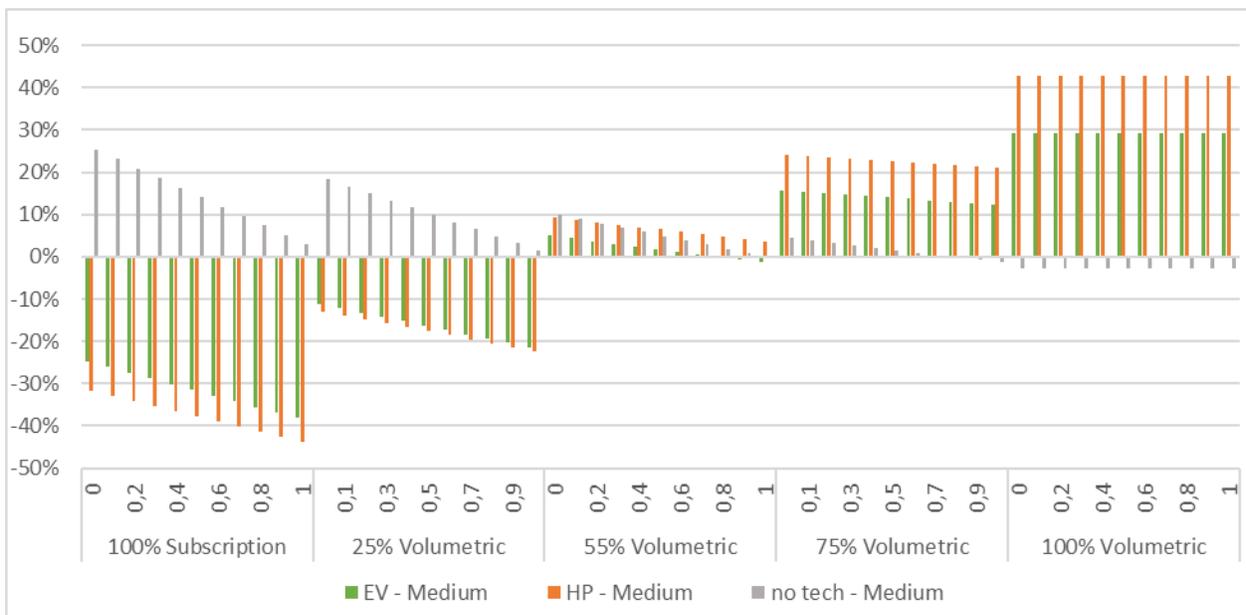

*Figure 7: Impacts of the redistribution factor for each tariff design and consumer group, with and without advanced equipment for medium financial status households*

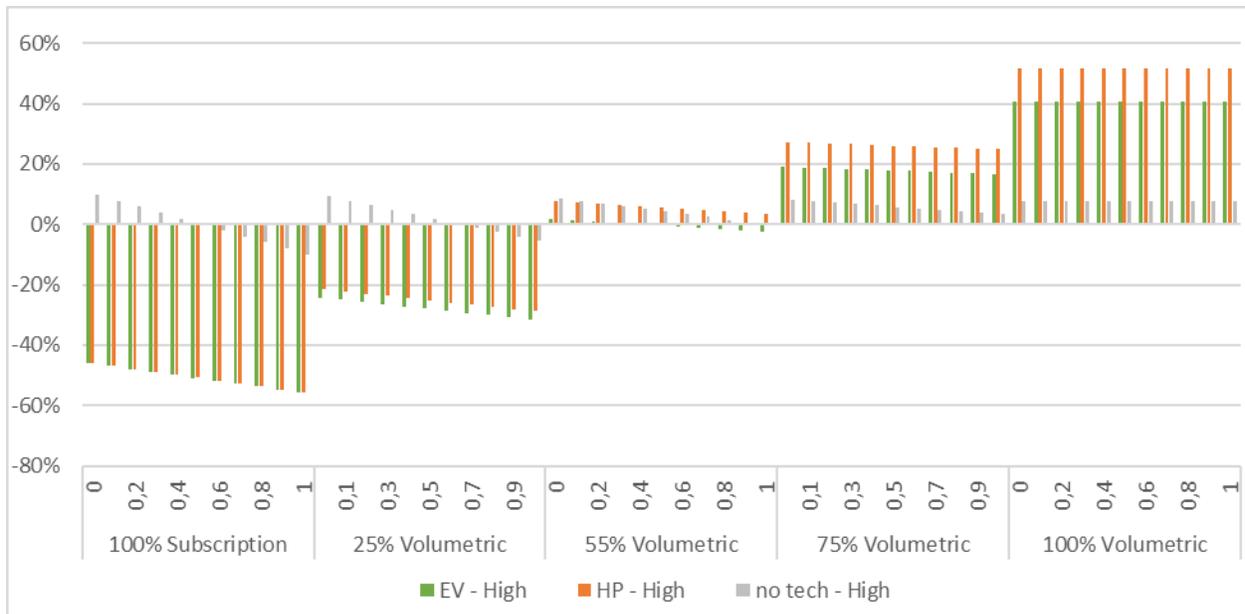

*Figure 8: Impacts of the redistribution factor for each tariff design and consumer group, with and without advanced equipment for high financial status households*

In addition to these expected results, the findings point to three original redistributive effects. First, there is a tipping point at the 55% volumetric tariff above which, and with some exceptions, the contribution of the high and medium financial status groups to grid cost recovery is always higher than in the base case (Figure 7, Figure 8), while keeping on benefiting to all low financial status households. At the 75% volumetric tariffs, only the low financial status households without HP benefit from the factor (Figure 6).

Second, small redistribution factors significantly impact low financial status households without excessively damaging the situation of some subsidizing groups. In subscription-heavy tariffs (from 55% volumetric to 100% subscription-based), a redistribution factor of 0.9 to 0.6 is sufficient to entirely alleviate the over cost paid by almost all low financial status households without incurring over cost to the high financial status group. A factor of 0.9 and 0.8 in the 55% volumetric setup has no negative impact on all the subsidizing groups without EV and HP and entirely alleviates the over cost for low financial status households. This indicates that a balance exists between tariff design, redistribution factor, no or minimal loss for the subsidizing groups and benefit for the subsidized group.

Lastly, the redistributive policy affects the most medium financial status households without HP or EV when combined with a subscription-heavy tariff design. Under a 100% subscription tariff, the average network bill of the medium financial status group increases by 24%, while the high financial status group only experiences an increase in its total bill of less than 2% (group without technologies).

## 7. Discussion

The electrification of heating and transportation may change the costs imposed on grid users. Needing cost-reflective network cost recovery, many jurisdictions consider using higher fixed costs that reflect higher network grid use. These fixed subscription charges can give clear incentives for some technology adoption while ensuring stable income to the grid operator. We see a clear shift in the budgets paid by the different household types as the relative share of the volumetric component in the tariff decreases. As the volumetric part rises above 55%, a larger grid cost share is transferred to EV and HP owners. These costs are borne by the larger electricity consumption of HP owners and then by EV owners. Thus, high volumetric tariffs create substantial additional costs for these technologies, potentially limiting future investment.

If the transfer to fixed charges supports electrification, it also has disadvantages. Most low financial status households do not or only marginally benefit from a subscription-only tariff. Especially, low financial status households without HP or EV subsidize the network connection of other households with better financial means and high-intensity equipment. The groups with medium and high financial status owning an EV or HP save 46% to 57% of their grid bill under the subscription tariff. This unfairness is a significant concern for using higher fixed charges in the context of electrification.

Cost redistribution can partially remediate inequities in grid cost payments. This tool benefits low financial status households and may support heating electrification, even if calibrated at a low level and therefore with low impact on other users. However, a high redistribution factor primarily affects medium financial status users. This outcome should alert us that these households, which are mostly middle-class, are likely to be the most affected by such a redistributive policy if it is poorly calibrated.

Finally, we show that a good tariff is possible. Here, "good" refers to the possibility of sending time-differentiated signals with a solid cost recovery base for the utility, which does not hinder the development of EVs and HPs and limits the transfer of grid costs to the most vulnerable households. We show that a subscription-heavy tariff associated with a ToU rate, and a low redistribution factor tackle all the above. In the 25% volumetric tariff, subsidizing the low financial status households by 10% (factor 0.9) alleviates all additional costs compared to the based case while only increasing the medium financial status group's bill by 3%.

## 8. Conclusion and policy implications

In this paper, we reported on equity analysis in grid tariff design under the growing adoption of HPs and EVs. We used a uniquely large dataset, both in the number of households and data for each household, from Denmark in 2017 to consider cross-subsidies due to changing tariffs from the current default tariff. These tariff changes were two-fold: first, the fixed subscription rate was changed to higher or lower levels of the full tariff, and second, a peak pricing rate was added. Besides, we tested an original redistribution mechanism to transfer part to all of the grid revenue from low to higher financial status households.

The rising attention to fairness, energy poverty, and access to clean energy stresses a potential rising conflict between the role of public policy and electricity rate making, at least in Europe. Today, DSOs can take network operation into account in setting tariffs, but setting "fair" tariffs, especially across income groups as motivated in this paper, is neither the aim of the DSO nor the regulator. Quite the opposite, the independent regulator and the EU legislation ensure that tariffs are not used to do politics. Currently, many policy measures are being introduced to shield energy consumers against energy price increase, sometimes without much targeting. We show that embedding a redistribution instrument to a grid tariff can strike a balance between the economic mandate of grid tariffs and public goals. This instrument can be a viable option to limit tariffs' worst inequities.

The results show the complex inter-relations between EV and HP ownership, financial status, and the design of the tariffs. When 10-20% of subscription costs are redistributed, low financial status households are protected against the larger fluctuations in bills arising from higher subscription fees. At this level of redistribution, there is a minor impact on household bills for households with better financial status, and the equity objective is better met. Simultaneously, a fixed subscription charge does not impact the viability of EVs and HPs against alternatives, thus meeting the climate change mitigation objective. However, our results also point to the observation that the variations in tariffs with a redistribution factor penalize the medium financial status group without EV/HP to a greater extent while leaving the better-off households with little impact.

Subscription-heavy tariffs support electrification, allowing households with HP or EV to save on their annual grid bill. On the other hand, volumetric-heavy tariffs protect low financial status grid users without advanced equipment from increasing network expenses. Hence, there is a trade-off between two policy goals: encouraging electrification and equity and energy poverty. The peak cost component makes comparatively minor changes to cross-subsidies between households. We set a peak price that resulted in 20% of costs being covered during 5% of highest demand hours. This resulted in a peak price about 4.6 times higher than the base volumetric rate. A higher peak price might induce further cross-subsidies from EV and HP owners, who drive peak demand hours, to non-owners.

*Limitations and future research*

Our results assume no (short-term) price elasticity of demand. Generally, the residential electricity sector has been considered to have negligible responsiveness to short-term changes in price (Labandeira, Labeaga, and López-Otero 2017; Burke and Abayasekara 2018; Simshauser 2016). However, a future with more communication and control infrastructure and more controllable electric devices (such as EVs) would increase short time price elasticity. Flexibility should then be considered in light of the price difference between time blocks. Past studies give insights on such levels of flexibility associated with a ToU tariff structure, albeit the scope of the studies may vary substantially. (A Faruqui and Palmer 2012) summarizes a multi-year effort to review and synthesize in detail the impacts of time-based trials in the U.S. The trials concerned time-based tariffs modelled on the wholesale prices and including network costs in the context of bundled activities and showed an average elasticity of substitution (from peak to off-peak consumption) of 0.5 with a ToU without advanced equipment. This elasticity may in some circumstances result in fairness concerns, as these changes depend on financial investments in technology and would thus not be equally distributed among society. Future research could focus on the impact of short-term elasticity on equity in a scenario with HPs and EVs.

**Acknowledgements**: The authors warmly thank Frederik Roose Øvlisen from the Danish utilities regulatory agency, FORSYNINGSTILSYNET, for his insights and valuable comments on this article.

**Funding**. This article was partly funded by the FlexSUS project (nbr. 91352) that received funding in the framework of the joint programming initiative ERA-Net RegSus, with support from the European Union's Horizon 2020 research and innovation programme under grant agreement No 775970.

# Appendices

*Appendix 1: Relative share of the household categories in the electricity demand and population.*

| | | Occupancy | Relative share in total consumption | Relative share in the population |
|---|---|---|---|---|
| **User category considered "low financial status."** | | | **10,32%** | **17,96%** |
| Low-income group with small area dwelling (house and apartment) | | P5+ | 0,07% | 0,05% |
| | | P3+ | 0,45% | 0,38% |
| | | P2 | 1,83% | 2,59% |
| | | P1 | 7,80% | 14,85% |
| | With heat pump | P1 & P2 | 0,17% | 0,10% |
| **User category considered as "medium financial status."** | | | **61,91%** | **63,82%** |
| High-income group with medium and small dwelling area (house and apartment) without advanced equipment | | All | 20,51% | 15,92% |
| | With heat pump | All | 0,22% | 0,07% |
| | With EV | All | 0,04% | 0,02% |
| Medium-income group with all dwelling area (house and apartment) without advanced equipment | | All | 29,87% | 31,90% |
| | With heat pump | All | 0,60% | 0,23% |
| Low-income group with medium area dwellings (house and apartment) without advanced equipment | | All | 10,45% | 15,58% |
| | With heat pump | All | 0,23% | 0,11% |
| **User category considered as "high financial status."** | | | **27,77%** | **18,22%** |
| High-income group with large area dwelling (house and apartment) without advanced equipment | | P5+ | 4,07% | 2,12% |
| | | P3+ | 13,79% | 8,75% |
| | | P2 | 8,27% | 6,31% |
| | | P1 | 0,77% | 0,80% |
| | With EV | P5+;EV1 | 0,08% | 0,02% |
| | | P3+;EV1 | 0,17% | 0,05% |
| | With heat pump | P5+;HP1 | 0,10% | 0,03% |
| | | P3+;HP1 | 0,28% | 0,08% |

|  |  |  | P2;HP1 | 0,25% | 0,07% |

*Appendix 2: Average grid bill per household category under factor 1 and 0*

|  |  | Base case 2017 | Factor 1 | | | | | Factor 0 | | | | |
|---|---|---|---|---|---|---|---|---|---|---|---|---|
|  |  |  | 100% Subs | 25% Vol | 55% Vol | 75% Vol | 100% Vol | 100% Subs | 25% Vol | 55% Vol | 75% Vol | 100% Vol |
| Low | No Tech | 873 | 937,6 | 907,7 | 871,8 | 847,9 | 818 | 0 | 204,5 | 449,9 | 613,5 | 818 |
| Low | HP | 1336 | 937,6 | 1143 | 1389 | 1553 | 1758 | 0 | 439,5 | 966,9 | 1318 | 1758 |
| Medium | No Tech | 973 | 937,6 | 953,4 | 972,4 | 985 | 1001 | 1143 | 1107 | 1065 | 1036 | 1001 |
| Medium | HP | 1738 | 937,6 | 1331 | 1802 | 2117 | 2510 | 1143 | 1485 | 1895 | 2168 | 2510 |
| Medium | EV | 1517 | 937,6 | 1193 | 1500 | 1704 | 1960 | 1143 | 1347 | 1592 | 1756 | 1960 |
| High | No Tech | 1083 | 937,6 | 1003 | 1081 | 1133 | 1198 | 1143 | 1157 | 1173 | 1184 | 1198 |
| High | HP | 2166 | 937,6 | 1529 | 2239 | 2712 | 3303 | 1143 | 1683 | 2331 | 2763 | 3303 |
| High | EV | 2118 | 937,6 | 1449 | 2063 | 2472 | 2984 | 1143 | 1603 | 2155 | 2523 | 2984 |

*Appendix 3: Relative share of each tariff cost component in the yearly average bill per consumer category under factor 1 and 0*

|  | Household type | RF | 1 | | | | | 0 | | | | |
|---|---|---|---|---|---|---|---|---|---|---|---|---|
|  |  | Vol% | 0% | 25% | 55% | 75% | 100% | 0% | 25% | 55% | 75% | 100% |
| Subs part | Low | No Tech | 100% | 78% | 51% | 31% | 0% | 0% | 0% | 0% | 0% | 0% |
| Subs part | Low | HP | 100% | 62% | 31% | 15% | 0% | 0% | 0% | 0% | 0% | 0% |
| Subs part | Medium | No Tech | 100% | 75% | 46% | 27% | 0% | 100% | 78% | 51% | 31% | 0% |
| Subs part | Medium | HP | 100% | 54% | 24% | 12% | 0% | 100% | 58% | 28% | 14% | 0% |
| Subs part | Medium | EV | 100% | 59% | 28% | 14% | 0% | 100% | 64% | 32% | 16% | 0% |
| Subs part | High | No Tech | 100% | 71% | 41% | 22% | 0% | 100% | 75% | 45% | 26% | 0% |
| Subs part | High | HP | 100% | 46% | 19% | 9% | 0% | 100% | 51% | 23% | 11% | 0% |
| Subs part | High | EV | 100% | 49% | 20% | 9% | 0% | 100% | 53% | 24% | 11% | 0% |
| Vol- Off-peak part | Low | No Tech | 0% | 16% | 37% | 52% | 75% | 0% | 75% | 75% | 75% | 75% |
| Vol - Off-peak part | Low | HP | 0% | 26% | 48% | 58% | 69% | 0% | 69% | 69% | 69% | 69% |
| Vol - Off-peak part | Medium | No Tech | 0% | 19% | 40% | 55% | 75% | 0% | 16% | 37% | 52% | 75% |

| | | | | | | | | | | | |
|---|---|---|---|---|---|---|---|---|---|---|---|
| Vol - Off-peak part | Medium | HP | 0% | 32% | 53% | 61% | 69% | 0% | 29% | 50% | 60% | 69% |
| Vol - Off-peak part | Medium | EV | 0% | 31% | 55% | 66% | 76% | 0% | 28% | 51% | 64% | 76% |
| Vol - Off-peak part | High | No Tech | 0% | 22% | 44% | 58% | 75% | 0% | 19% | 41% | 56% | 75% |
| Vol - Off-peak part | High | HP | 0% | 38% | 57% | 64% | 70% | 0% | 34% | 54% | 63% | 70% |
| Vol - Off-peak part | High | EV | 0% | 40% | 62% | 70% | 78% | 0% | 36% | 59% | 69% | 78% |
| Vol - Peak part | Low | No Tech | 0% | 6% | 12% | 17% | 25% | 0% | 25% | 25% | 25% | 25% |
| Vol - Peak part | Low | HP | 0% | 12% | 22% | 26% | 31% | 0% | 31% | 31% | 31% | 31% |
| Vol - Peak part | Medium | No Tech | 0% | 6% | 14% | 18% | 25% | 0% | 6% | 12% | 17% | 25% |
| Vol - Peak part | Medium | HP | 0% | 14% | 23% | 27% | 31% | 0% | 13% | 22% | 26% | 31% |
| Vol - Peak part | Medium | EV | 0% | 10% | 17% | 21% | 24% | 0% | 9% | 16% | 20% | 24% |
| Vol - Peak part | High | No Tech | 0% | 7% | 15% | 20% | 25% | 0% | 6% | 14% | 19% | 25% |
| Vol - Peak part | High | HP | 0% | 16% | 24% | 27% | 30% | 0% | 14% | 23% | 27% | 30% |
| Vol - Peak part | High | EV | 0% | 11% | 18% | 20% | 22% | 0% | 10% | 17% | 20% | 22% |

*RF: Redistribution Factor*

*Appendix 4: Detailed impact of the redistribution factor for each tariff scenario and consumer group with and without EV and HP. The values indicate the difference compared to base case.*

| Tariff design | R.F. | High | | | Medium | | | Low | |
|---|---|---|---|---|---|---|---|---|---|
| | | EV | HP | No tech | EV | HP | No tech | HP | No tech |
| 100% subscription | 0 | -45,98% | -45,93% | 9,80% | -24,66% | -31,61% | 25,39% | -100,00% | -100,00% |
| | 0,1 | -46,95% | -46,90% | 7,83% | -26,02% | -32,84% | 23,14% | -92,90% | -88,68% |
| | 0,2 | -47,92% | -47,87% | 5,86% | -27,37% | -34,06% | 20,89% | -85,81% | -77,36% |
| | 0,3 | -48,89% | -48,84% | 3,88% | -28,72% | -35,29% | 18,63% | -78,71% | -66,04% |
| | 0,4 | -49,86% | -49,81% | 1,91% | -30,07% | -36,52% | 16,38% | -71,61% | -54,72% |
| | 0,5 | -50,83% | -50,78% | -0,06% | -31,43% | -37,75% | 14,13% | -64,52% | -43,40% |
| | 0,6 | -51,80% | -51,75% | -2,03% | -32,78% | -38,98% | 11,88% | -57,42% | -32,09% |
| | 0,7 | -52,77% | -52,73% | -4,00% | -34,13% | -40,21% | 9,63% | -50,32% | -20,77% |
| | 0,8 | -53,74% | -53,70% | -5,98% | -35,49% | -41,43% | 7,38% | -43,22% | -9,45% |
| | 0,9 | -54,71% | -54,67% | -7,95% | -36,84% | -42,66% | 5,12% | -36,13% | 1,87% |
| | 1 | -55,68% | -55,64% | -9,92% | -38,19% | -43,89% | 2,87% | -29,03% | 13,19% |
| 25% volumetric | 0 | -24,27% | -21,53% | 9,29% | -11,20% | -13,04% | 18,38% | -67,27% | -77,78% |
| | 0,1 | -25,00% | -22,26% | 7,81% | -12,21% | -13,96% | 16,69% | -61,95% | -69,29% |
| | 0,2 | -25,73% | -22,98% | 6,33% | -13,23% | -14,88% | 15,00% | -56,63% | -60,80% |
| | 0,3 | -26,46% | -23,71% | 4,85% | -14,24% | -15,80% | 13,32% | -51,30% | -52,31% |
| | 0,4 | -27,18% | -24,44% | 3,37% | -15,26% | -16,72% | 11,63% | -45,98% | -43,82% |
| | 0,5 | -27,91% | -25,17% | 1,90% | -16,27% | -17,64% | 9,94% | -40,66% | -35,33% |
| | 0,6 | -28,64% | -25,90% | 0,42% | -17,29% | -18,57% | 8,25% | -35,34% | -26,84% |
| | 0,7 | -29,37% | -26,63% | -1,06% | -18,30% | -19,49% | 6,56% | -30,01% | -18,35% |
| | 0,8 | -30,09% | -27,35% | -2,54% | -19,32% | -20,41% | 4,87% | -24,69% | -9,86% |
| | 0,9 | -30,82% | -28,08% | -4,02% | -20,33% | -21,33% | 3,18% | -19,37% | -1,37% |
| | 1 | -31,55% | -28,81% | -5,50% | -21,35% | -22,25% | 1,49% | -14,04% | 7,12% |

|  | RF | | | | | | | | |
|---|---|---|---|---|---|---|---|---|---|
| 55% volumetric | 0 | 1,77% | 7,75% | 8,68% | 4,96% | 9,24% | 9,97% | -28,00% | -51,11% |
| | 0,1 | 1,33% | 7,32% | 7,79% | 4,35% | 8,69% | 8,96% | -24,80% | -46,01% |
| | 0,2 | 0,90% | 6,88% | 6,90% | 3,74% | 8,14% | 7,95% | -21,61% | -40,92% |
| | 0,3 | 0,46% | 6,44% | 6,02% | 3,13% | 7,59% | 6,93% | -18,42% | -35,83% |
| | 0,4 | 0,02% | 6,00% | 5,13% | 2,53% | 7,03% | 5,92% | -15,22% | -30,73% |
| | 0,5 | -0,41% | 5,57% | 4,24% | 1,92% | 6,48% | 4,91% | -12,03% | -25,64% |
| | 0,6 | -0,85% | 5,13% | 3,35% | 1,31% | 5,93% | 3,89% | -8,84% | -20,54% |
| | 0,7 | -1,29% | 4,69% | 2,47% | 0,70% | 5,38% | 2,88% | -5,64% | -15,45% |
| | 0,8 | -1,72% | 4,26% | 1,58% | 0,09% | 4,82% | 1,87% | -2,45% | -10,36% |
| | 0,9 | -2,16% | 3,82% | 0,69% | -0,52% | 4,27% | 0,85% | 0,75% | -5,26% |
| | 1 | -2,60% | 3,38% | -0,20% | -1,13% | 3,72% | -0,16% | 3,94% | -0,17% |
| 75% volumetric | 0 | 19,13% | 27,27% | 8,27% | 15,73% | 24,10% | 4,37% | -1,81% | -33,33% |
| | 0,1 | 18,89% | 27,03% | 7,78% | 15,40% | 23,79% | 3,80% | -0,04% | -30,50% |
| | 0,2 | 18,65% | 26,79% | 7,28% | 15,06% | 23,49% | 3,24% | 1,73% | -27,67% |
| | 0,3 | 18,40% | 26,54% | 6,79% | 14,72% | 23,18% | 2,68% | 3,51% | -24,84% |
| | 0,4 | 18,16% | 26,30% | 6,30% | 14,38% | 22,87% | 2,11% | 5,28% | -22,01% |
| | 0,5 | 17,92% | 26,06% | 5,81% | 14,04% | 22,56% | 1,55% | 7,06% | -19,18% |
| | 0,6 | 17,68% | 25,82% | 5,31% | 13,70% | 22,26% | 0,99% | 8,83% | -16,35% |
| | 0,7 | 17,43% | 25,57% | 4,82% | 13,37% | 21,95% | 0,43% | 10,61% | -13,52% |
| | 0,8 | 17,19% | 25,33% | 4,33% | 13,03% | 21,64% | -0,14% | 12,38% | -10,69% |
| | 0,9 | 16,95% | 25,09% | 3,83% | 12,69% | 21,34% | -0,70% | 14,15% | -7,86% |
| | 1 | 16,71% | 24,84% | 3,34% | 12,35% | 21,03% | -1,26% | 15,93% | -5,03% |
| 100% volumetric | 0 | 40,83% | 51,67% | 7,76% | 29,20% | 42,67% | -2,64% | 30,91% | -11,10% |
| | 0,1 | 40,83% | 51,67% | 7,76% | 29,20% | 42,67% | -2,64% | 30,91% | -11,10% |
| | 0,2 | 40,83% | 51,67% | 7,76% | 29,20% | 42,67% | -2,64% | 30,91% | -11,10% |
| | 0,3 | 40,83% | 51,67% | 7,76% | 29,20% | 42,67% | -2,64% | 30,91% | -11,10% |
| | 0,4 | 40,83% | 51,67% | 7,76% | 29,20% | 42,67% | -2,64% | 30,91% | -11,10% |
| | 0,5 | 40,83% | 51,67% | 7,76% | 29,20% | 42,67% | -2,64% | 30,91% | -11,10% |
| | 0,6 | 40,83% | 51,67% | 7,76% | 29,20% | 42,67% | -2,64% | 30,91% | -11,10% |
| | 0,7 | 40,83% | 51,67% | 7,76% | 29,20% | 42,67% | -2,64% | 30,91% | -11,10% |
| | 0,8 | 40,83% | 51,67% | 7,76% | 29,20% | 42,67% | -2,64% | 30,91% | -11,10% |
| | 0,9 | 40,83% | 51,67% | 7,76% | 29,20% | 42,67% | -2,64% | 30,91% | -11,10% |
| | 1 | 40,83% | 51,67% | 7,76% | 29,20% | 42,67% | -2,64% | 30,91% | -11,10% |

*RF: Redistribution Factor*

*Appendix 5: Aggregated grid bill paid per consumer category with and without advanced equipment in the base case (Million DKK)*

|  | EV | HP | No tech | Grand Total |
|---|---|---|---|---|
| **High** | 2307995,3 | 5835177,2 | 304295587,1 | 312438759,7 |
| **Medium** | 394434,3 | 10398486,7 | 870946674,0 | 881739595,0 |
| **Low** | 0,0 | 1933108,9 | 180997253,5 | 182930362,4 |
| **Total** | 2702429,6 | 18166772,8 | 1356239515,0 | 1377108717,0 |